# Femtosecond drift photocurrents generated by an inversely designed plasmonic antenna

Ye Mou[1], Xingyu Yang[1], Marlo Vega[2,3,4] Bruno Gallas[1], Jean-François Bryche[2,3], Alexandre Bouhelier[5] and Mathieu Mivelle[1,*]

[1]Sorbonne Université, CNRS, Institut des NanoSciences de Paris, INSP, F-75005 Paris, France.

[2]Laboratoire Nanotechnologies Nanosystèmes (LN2)-IRL3463, CNRS, Université de Sherbrooke, Université Grenoble Alpes, École Centrale de Lyon, INSA Lyon, Sherbrooke, J1K 0A5 Québec, Canada. [3]Institut Interdisciplinaire d'Innovation Technologique (3IT), Université de Sherbrooke, 3000 Boulevard de l'université, Sherbrooke, J1K OA5 Québec, Canada

[4]Université Paris-Saclay, Institut d'Optique Graduate School, CNRS, Laboratoire Charles Fabry, Palaiseau, France.

[5]Laboratoire Interdisciplinaire Carnot de Bourgogne, CNRS UMR 6303 Université de Bourgogne, 21000 Dijon, France.

*Corresponding author: mathieu.mivelle@sorbonne-universite.fr

ORCID:0000-0002-0648-7134

## Abstract

Photocurrents play a crucial role in various applications, including light detection, photovoltaics, and THz radiation generation. Despite the abundance of methods and materials for converting light into electrical signals, the use of metals in this context has been relatively limited. Nanostructures supporting surface plasmons in metals offer precise light manipulation and induce light-driven electron motion. Through inverse design optimization of a gold nanostructure, we demonstrate enhanced volumetric, unidirectional, intense, and ultrafast photocurrents via a magneto-optical process derived from the inverse Faraday effect. This is achieved through fine-tuning the amplitude, polarization, and their gradients in the local light field. The virtually instantaneous process allows dynamic photocurrent modulation by varying optical pulse duration, potentially yielding nanosources of intense, ultrafast, planar magnetic fields, and frequency-tunable THz emission. These findings opens avenues for ultrafast magnetic material manipulation and holds promise for nanoscale THz spectroscopy.

## Introduction

A photocurrent is the flow of electric charges generated when photons are absorbed by a material. This phenomenon plays a critical role in various groundbreaking fields like light detection, solar photovoltaics, lightwave electronics, and THz spectroscopy and imaging.[1] For physicists, a direct photocurrent holds significant importance as it serves as a unique tool to investigate various processes across a wide range of spatial and temporal scales, ranging from femtoseconds to milliseconds and from the mesoscale to the microscale. Hence, understanding photocurrent is crucial to unveil physical and chemical effects due to the intimate connection between photons and electrons.

Photocurrents have been instrumental in advancing the development of low-dimensional structures,[2] topological materials,[3] and the exploration of systems exhibiting symmetry breaking.[4,5] Their often ultrafast dynamics contribute to understand out-of-equilibrium transient states.[6,7] Broadly speaking, the conversion of light into an electrical signal can result from various phenomena. Some, like the p-n junction, are commonly found in everyday technologies, while others, such as current generated by photothermal gradients, are less explored in the literature. There are fascinating prospects involving photo-induced transport of charge carriers in metals. Examples include optical rectification of visible photons through plasmonic nanoscale rectennas[8-10] or the generation of photocurrents by structured light with variations in intensity, phase, and polarization.[11] What makes metals and plasmonics particularly appealing is their ability to fine-tune local optical fields while simultaneously generating electric currents on the same material platform, thereby eliminating the need for complex hetero-integration of materials.

Despite recent efforts to understand the interplay of phenomena in these systems,[12, 13] a comprehensive framework for manipulating direct photocurrents using plasmonics at the nanoscale is still largely missing. A strategy to engineer their spatial distribution, control their strength, and dictate their direction remains to be rigorously formulated theoretically, optimized through numerical simulations, and verified experimentally.

There are physical processes that has been generally overlooked for photocurrent generation: the effect of nonlinear forces. A ponderomotive force is an example of such processes enabling drift currents in the skin depth of a metal parallel to its surface. These drift currents have many advantages: they are relatively simple to model, their generation is almost instantaneous, and they last only as long as the light exposure, so they can be very short when generated by an optical pulse. Moreover, they can be manipulated at will by tuning the intensity of the optical field, its polarization, or its gradients.

Drift currents can be regarded as a type of optical rectification at the origin of the inverse Faraday effect in metals.[14-16] This is a nonlinear process promoting the magnetization of matter by an optical excitation only. Its use recently spurred the theoretical[17, 18] and experimental[20, 21] demonstration of very intense and very short magnetic fields at the nanoscale, with potential applications in data writing and processing at ultrafast time scales. In these studies, circular polarizations are used to move electrons in structures with a rotational symmetry. This is the principle of the inverse Faraday effect (IFE) described in 1960s. However, it was recently demonstrated that new physical effects can emerge from IFE by manipulating light at the nanoscale with plasmonic nanostructures. In particular, it was shown that IFE can be generated by linear polarization,[19] that it can be chiral,[20, 21] or even have its symmetry reversed.[22]. Our study builds upon this body of literature. We demonstrate that a plasmonic nanostructure specifically designed by inverse optimization manipulates light field at the nanoscale in such a way that a volumetric, unidirectional, intense, and ultra-short photocurrent emerges. We also demonstrate that this phenomenon is chiral, *i.e.,* occurring for only one polarization of light and that it can generate a magnetic field perpendicular to the direction of light propagation, which is impossible with a conventional inverse Faraday effect. Finally, by generating ultra-short, planar and unidirectional photocurrents, we show that this physical process opens the way to generate a linearly polarized and frequency-tunable THz nanosource.

## Results and discussions

In this work, we use the optical rectification phenomenon at the origin of IFE to generate ultrafast direct photocurrents at the nanoscale by a light-induced magnetization of matter.[14-16] In a metal, this magnetization is made possible by the nonlinear forces that an electromagnetic wave applies to the free electrons. The average value of these forces being non-zero, a direct current called drift current is generated, and in turn, forms a stationary magnetic field.

The theory underlying the generation of drift currents was first developed by the plasma community[23, 24] before being extrapolated to optics[25, 26] and nanophotonics.[17-21, 27-31] The following equation gives the expression of the drift currents in a metal:[18, 19]

$$\boldsymbol{J_d} = \frac{1}{2en} Re\left(\left(-\frac{\nabla \cdot (\sigma_\omega \boldsymbol{E})}{i\omega}\right) \cdot (\sigma_\omega \boldsymbol{E})^*\right) \quad (1)$$

$e$ is the charge of the electron ($e < 0$), $n$ is the charge density at rest, $\sigma_\omega$ is the dynamic conductivity of the metal and **E** is the optical electric field. The terms in bold represent vectorial values, and the star * denotes the conjugate complex.

As one can see in equation 1, drift currents are proportional to the optical electric field and its divergence. Therefore, by its ability to enhance electric fields and to create strong field gradients, plasmonics is a particularly well-suited technology to generate and manipulate in a control manner strong drift currents with the aim to create ultrafast, intense magnetic field pulses at the nanoscale.[18]

In IFE, it is conventionally understood that drift currents have an azimuthal symmetry generating a magnetization perpendicular to the plane in which the electrons spin, as recently demonstrated in a series of theoretical[17-19, 21, 28, 30] and experimental[29, 31] papers. However, as recognized by the plasma community, the nonlinear forces that light applies to the electron sea possess all kinds of symmetries. For instance, ponderomotive forces accelerating charges in a plasma hold radial symmetries.[32, 33]

Based on this concept, we designed a plasmonic nanostructure capable of generating a direct drift photocurrent that is intense, unidirectional in the plane of the object, independent of the excitation polarization, and ultrafast. The origin of these exceptional properties can be found in equation 1. As shown in this expression, the drift currents can only be created when the product **E.E**\* is not zero, *i.e.,* for an elliptical or circular polarization of the light. Therefore manipulating the local polarization of the light with an adequate design enables to create different ellipticities around the object and a control of the drift currents.

Specifically, we have inversely designed a plasmonic nanostructure in a 30 nm thick gold layer deposited on a glass substrate using a genetic algorithm (GA) based on topological optimization (Figure 1).

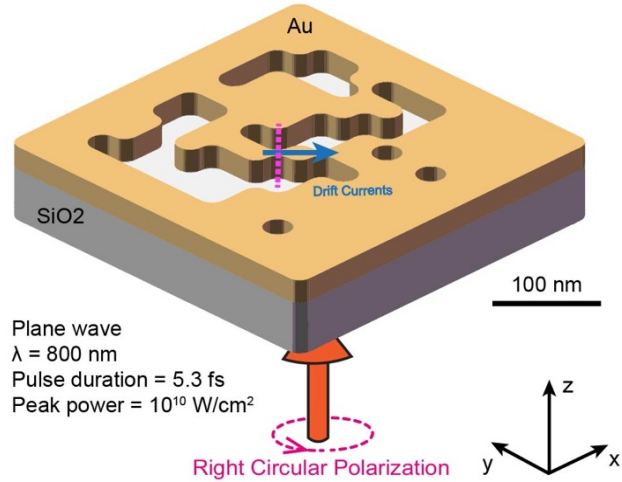

***Figure 1***. *Optimized structure and excitation conditions. Example of a GA-optimized structure, made in a 30 nm thick gold layer.* The 280 x 280 nm$^2$ nanostructure is based on a *two-dimensional array of 10x10 elements, each element consisting of metal or air with a size of 28 nm. These dimensions are chosen to facilitate the fabrication of this structure by e.g. nanoscale lithography techniques. For the same reason and to avoid non-physical effects that a numerical approach can generate locally, the corners of the nanostructure are rounded (see figure S1 of the Supporting Information). The optical excitation of the nanostructure is modeled by a pulsed plane wave emitted at a wavelength of 800 nm. The pulse is launched from the substrate side and is right circularly polarized. We set the pulse duration at 5.3 fs and the peak power of 10$^{10}$ W/cm$^2$, a value well below the damage threshold of the material.*[34, 35] *The blue arrow represents the ultrafast direct currents flowing in a single direction, and the pink dashed line indicates the position of a cross-section.*

In order to generate a strong unidirectional drift current in the plane of the metal layer, the GA optimization function consisted in maximizing the local stationary magnetic field created by the drift currents at the center of the structure and having a vectorial symmetry parallel to the gold surface. Figure S2 represents the evolutionary history of the nanostructures in connection with the generation of a strong magnetic field by IFE. After 87 generations the strength of the magnetic field reaches a plateau, and the structure is considered optimized. Figure 2a represents schematically the result of the optimization procedure. The interested part of the structure is certainly the bridge-like geometry at the center. Figure 2b shows the distribution and the increase of the optical electric field at the origin of the drift currents and **B** field in the center of this design (represented by a black square in figure 2a) in an XY plane taken at the middle of the metal thickness. From this electric field distribution and using Equation 1, we can then plot the distribution of drift currents in the structure as shown in Figure 2c. First, drift currents are maximal in a gold strip in the center of the structure but not necessarily where the electric field is the strongest. Importantly, the arrows representing the direction of drift are all oriented in the same direction in the XY plane, demonstrating the possibility of generating, in a controlled manner, straight and unidirectional DC currents by optical excitation. Note that other physical processes may be responsible for generating DC currents, one of which is related to thermal effects, for instance. Although the message of this article is not to compare all these processes, a study of temperature generation in our optimized nano-system for our optical excitation characteristics is described in Supporting Information (Figure S3) based on two-temperature model.[36] As described in this supporting figure, the lattice temperature remains extremely modest and rather homogenous in the gold strip, implying a minimal contribution of this physical process to DC current generation. Also, the electronic temperature for the peak powers and pulse durations considered here are relatively low (Figure S4).

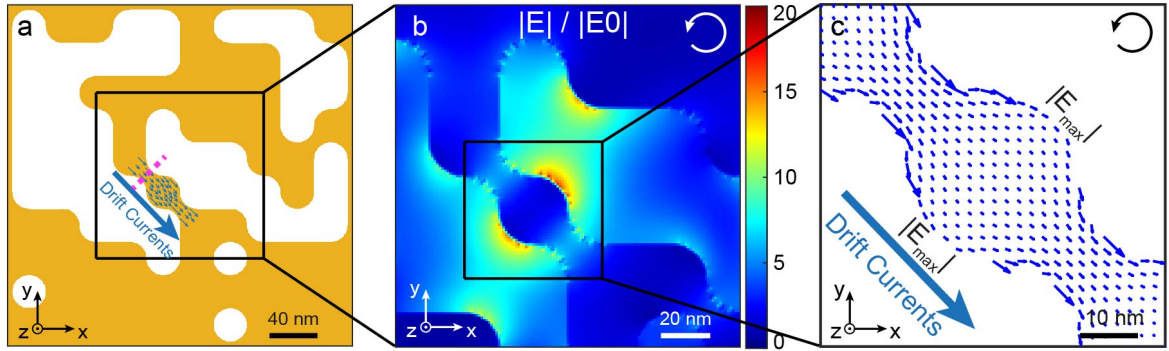

*Figure 2. Optical response of the optimized plasmonic nanostructure and associated drift currents. a) Schematic, in an XY plane, of the GA-optimized structure. b) Spatial distribution of the electric field enhancement at the Z-center of the structure shown in a) for a right circular polarization of excitation. c) Spatial distribution of the drift currents generated at the surface of the structure by the electric field distribution shown in b). The length of the arrows refers to the amplitude of the drift currents.*

One way to visualize the vectorial orientation of the DC currents in the metal is to image the resulting vectorial distribution of the magnetic field (**B**). For this purpose, Figure 3 represents the distribution of the three components of the magnetic field created by IFE in the same plane as Figure 2b,c. In particular, Figure 3a displays the amplitude of the total **B**-field and Figures 3b-d, respectively, the **B**-field components oriented along X, Y, and Z. As one can see, on the one hand, the fields along X and Y are both negative. However, on the other hand, the one along Z is positive on one side of the gold strip and negative on the other. This distribution of the magnetic field components around the thin gold strip is therefore characteristic of a current flowing unidirectionally in the direction shown by the arrow in Figure 2c (see also figure S5). Moreover, we note that in past studies, the orientation of the magnetic fields generated by optical rectification through IFE have all been collinear with the direction of light propagation.[17, 18, 28] Here, by manipulating the light at the nanoscale, we demonstrate the creation of an azimuthally polarized magnetic field around the gold nanostrip (Figures 3b,c, and S5). This allows for generating a magnetic field above the gold layer parallel to the nanostructure's surface, and thus perpendicular to the direction of light propagation. This result has significant implications for applying IFE to manipulate magnetic processes at the nano[37] and potentially ultrafast[38, 39] scales since many magnetic layers have domains orientation parallel to the sample plane (i.e., XY here).

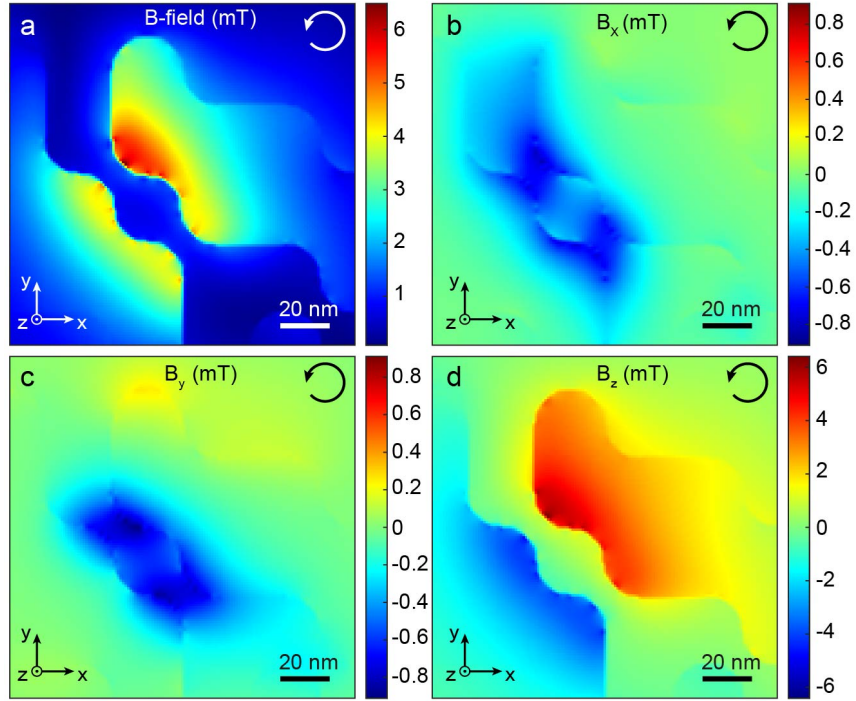

*Figure 3. Magnetic response of the optimized plasmonic nanostructure under a right circular polarization of excitation. a) Spatial distribution of the amplitude of the total magnetic field generated at the Z-center of the structure represented in Figure 2a. Decomposition of the spatial distribution of the magnetic field a) into the components oriented along b) X, c) Y, and d) Z.*

To further understand why this plasmonic structure creates a unidirectional DC current, we plot in Figure 4a the distribution of the Z-oriented component of the optical spin density in the same plane as Figure 2c, which is a parameter related to light polarization. The spin density **s**,[40-42] described by the following equation:

$$\bm{s} = \frac{1}{|E_0|^2} Im(\bm{E}^* \times \bm{E}) \quad (2)$$

is a vectorial quantity describing the polarization state of the light. In our coordinate system, negative and positive spin densities correspond respectively to left elliptical and right elliptical polarization; and s = 0 is equivalent to a linear polarization. Now, as highlighted above, drift currents can emerge in the presence of an elliptical polarization. From equation 1, a left elliptical polarization creates a drift current in one direction, while a right elliptical polarization generates drift currents in the opposite direction. With this understanding, the spin density distribution shown in Figure 4a sheds light on the vectorial distribution of DC currents shown in Figure 2c. On each side of the gold strip, where the currents are the strongest, we concomitantly observe the strongest spin densities of opposite signs. By symmetry, the generated drift currents are flowing in the same direction and are oriented in the same way as the red and blue arrows representing the local helicity of the light. These hot spots of so-called super circular light,[19, 21] a concept similar to super chiral light related to elliptical polarization of enhanced optical fields[43], are thus at the origin of the strong currents and their straight and unidirectional orientation. The spin density distribution inducing this physical effect is the result of constructive and destructive interferences of the light in the plasmonic structure, interferences that the evolutionary approach of the GA has selected.

Because the nanostructure is excited by an ultrashort optical pulse, we anticipate that the resulting DC current to carry also an ultrafast dynamics. Figure 4b confirms this hypothesis. This figure represents the instantaneous intensity of the electric field as a function of time at a point on the surface of the structure, symbolized by a pink star in Figure 4a and for an optical pulse of 5.3 fs. In the same figure is shown the total drift current generated by this pulse and passing through a section of the gold strip represented by the pink dotted line in Figure 4a (Y'Z section in the X'Y'Z coordinate system rotated from the XYZ coordinate system by an angle 45° around Z, as shown in figure 4a). Clearly, an ultrashort photocurrent transient emerges across the section of the gold strip and reaches up to 400 mA with the optical power used in the calculation. The duration of this transient is directly related to the duration of the optical pulse, since when there is no more energy to set the electrons in motion, they slow down immediately. Interestingly, this decay time can be observed in the tiny oscillations of the drift current, which occur when the instantaneous energy in the plasmonic nanostructure is 0 (taking into account a phase shift).

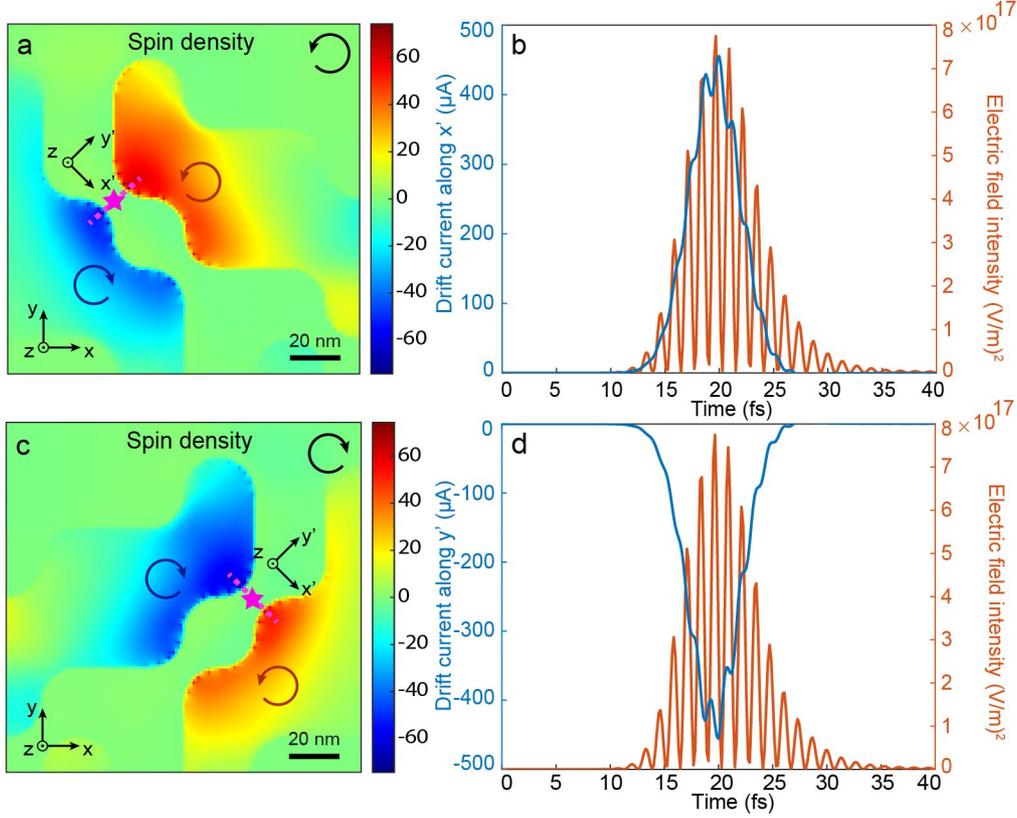

*Figure 4. Spin density distribution and temporal responses. a) Local spin density at the Z-center of the optimized plasmonic nanostructure shown in Figure 2a for a right circular polarized excitation. b) Temporal response of the drift currents flowing along X' through the section shown in a) as a dashed line and the electric field intensity at the pink star point shown in a). c) Spatial distribution of the spin density in the same plane and for the mirror structure shown in a) excited by a left circular polarization. d) Temporal response of the drift currents and the electric field intensity, respectively, through the section symbolized by a dotted line and the pink star shown in c).*

To further demonstrate the potentiality of controlling IFE-induced drift current at the nanoscale, an additional optimization condition was included in the definition of this nanostructure, consisting in making IFE chiral. During the selection process by the GA, the structures were selected to generate a magnetic field only for one helicity of the light, *i.e.* the right circular polarization. Therefore, when a left circular polarization illuminates the nanostructure, the drift currents and the associated stationary magnetic field are zero (Figures S6 and S7). Reciprocally, the mirror structure of Figure 2a can generate a symmetric spin density only when excited by a left circular polarization (Figure 4c) which results in an ultrafast direct current of perpendicular direction in the X'Y'Z coordinate system (Figures 4d and S8). These results are remarkable since this nanostructured landscape enables generating an ultrashort unidirectional current pulse even for a unpolarized excitation. Due to its chirality property, the GA-optimized structure naturally filters the polarizations from the unpolarized incoming field producing the direct current (principle of chirality).[44] Therefore, our algorithmic optimization approach isolates plasmonic nanostructures generating ultrafast photoinduced currents in one direction for a specific excitation polarization and in another direction for an alternative polarization condition.

Based on these findings and drawing inspiration from the potential applications of ultrafast photocurrents for THz generation, we assess the suitability of the transient drift currents to act as a tunable THz nanosource. To achieve this, we excite the structure depicted in Figure 2a with femtosecond (fs) pulses of varying durations and observe temporal responses of drift photocurrents. Our numerical simulations are displayed in Figure 5a. Because of the ultrafast current dynamics already highlighted above (see Figure 4), an increase of the pulse duration leads to a longer current transient. Importantly, since the peak power of the optical pulses is kept identical, differences in photocurrent intensity are primarily attributed to the more effective power build-up of the structure at longer pulses.[18]

From the temporal responses calculated in Figure 5a, we can extract radiation generated at various frequencies by performing a Fourier transform of the time derivative of the currents. Figure 5b displays these Fourier transforms. The spectra reveal two frequency components associated with each photocurrent time response: one at low frequencies and another centered at 749.5 THz. The first component, displayed at higher resolution in Figure 5c, is the THz radiation produced by the current impulse. The second component corresponds to the second harmonic generation resulting from the second-order nonlinear process described by equation 1.[45]

Subsequently, we plot in Figure 5c the peak THz frequencies generated by the optimized nanostructure as a function of optical pulse durations based on the Fourier transforms. As expected, these frequencies exhibit an inverse relationship with pulse lengths, ranging from 70 THz to approximately 4 THz for the selected pulse durations. An appealing feature of current transients produced by IFE is that they not only creates a nanoscale, tunable source of THz radiation but also impart linear polarization to that radiation, despite the circular excitation of the incident wave. This is made possible by the unidirectionality of the photocurrents drifting in the gold plane,

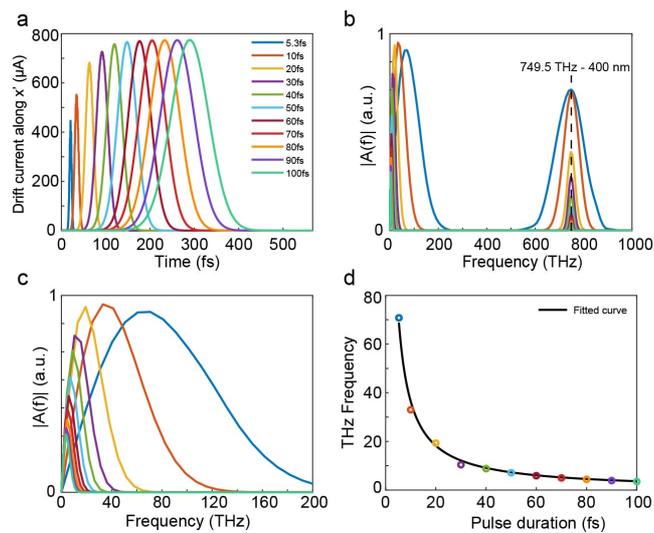

*Figure 5. Generation of tunable THz radiation. a) Time response of the drift photocurrents as a function of the duration of the light pulse exciting the optimized nanostructure described in Figure 2a. b) Fourier transform of the time responses shown in a) over the frequency range from 0 to 1000 THz. c) Fourier transform of the time responses shown in a) over the frequency range from 0 to 200 THz. d) Distribution of the THz frequencies generated as a function of the pulse duration used.*

In conclusion, we demonstrate that by manipulating the optical fields and field gradients in a plasmonic nanostructure optimized by an inverse design algorithm, unidirectional and direct photocurrents in the metal plane are generated. This is made possible by influencing the local polarization of light around the photonic nanostructure. Also, thanks to the ability to assign several optimization objectives to the genetic algorithm, the optimized nanostructure creates a direct photocurrent only for one helicity, here the right circular polarization. This chirality imparted to physical response enables generation of these currents even for unpolarized light. We then confirm that a direct current of the opposite direction is generated in the mirror structure for a left circular polarization, implying that a control of the current flow is possible only by changing the polarization incident on the structure. Also and importantly, we demonstrate that through the inverse Faraday effect, unidirectional current pulses in the plane of the metal are generated at ultrafast timescales. These currents can be harnessed to produce ultrafast magnetic fields with unexpected symmetry, defying conventional expectations associated with the inverse Faraday effect.

In addition, the ultrafast drift photocurrents generated in this way offer the potential for creating a versatile and tunable nanosource for linearly polarized THz radiation. Hence, these findings pave the way to ultrafast in-plane manipulation of magnetic domains and the possibility of conducting ultra-localized THz spectroscopy at the nanoscale.

**Acknowledgements**

The authors declare no conflict of interest.

We acknowledge the financial support from the Agence national de la Recherche (ANR-20-CE09-0031-01, ANR-23-ERCC-0005), from the Institut de Physique du CNRS (Tremplin@INP 2020) and the China Scholarship Council. This work has been partially funded by the French Agence Nationale de la Recherche (ISITE-BFC ANR-15-IDEX-0003), the EIPHI Graduate School (ANR-17-EURE-0002) and the European Union through the PO FEDER-FSE Bourgogne Franche-Comté 2021/2027 programs.

LN2 is a joint International Research Laboratory (IRL 3463) funded and co-operated in Canada by Université de Sherbrooke (UdeS) and in France by CNRS as well as ECL, INSA Lyon, and Université Grenoble Alpes (UGA). It is also supported by the Fonds de Recherche du Québec Nature et Technologie (FRQNT). This work was supported by the Natural Sciences and Engineering Research Council of Canada (NSERC) with Discovery Grants for JF Bryche. A doctoral scholarship from IDEX Paris-Saclay supports M. Vega. This work is supported by the "ADI" project funded by the IDEX Paris-Saclay, ANR-11-IDEX-0003-02.

Table of Content

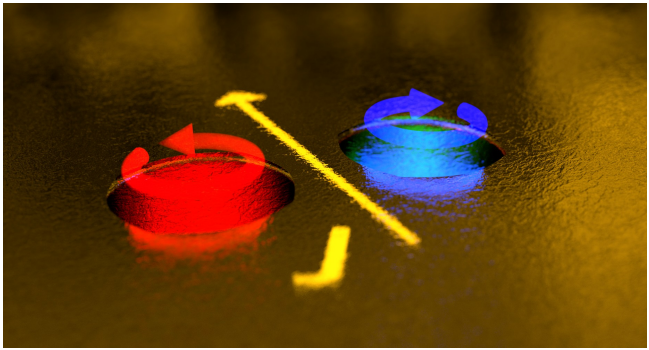

# Femtosecond drift photocurrents generated by an inversely designed plasmonic antenna

Ye Mou[1], Xingyu Yang[1], Marlo Vega[2,3,4] Bruno Gallas[1], Jean-François Bryche[2,3], Alexandre Bouhelier[5] and Mathieu Mivelle[1,*]

[1]Sorbonne Université, CNRS, Institut des NanoSciences de Paris, INSP, F-75005 Paris, France.

[2]Laboratoire Nanotechnologies Nanosystèmes (LN2)-IRL3463, CNRS, Université de Sherbrooke, Université Grenoble Alpes, École Centrale de Lyon, INSA Lyon, Sherbrooke, J1K 0A5 Québec, Canada.[3]Institut Interdisciplinaire d'Innovation Technologique (3IT), Université de Sherbrooke, 3000 Boulevard de l'université, Sherbrooke, J1K OA5 Québec, Canada

[4]Université Paris-Saclay, Institut d'Optique Graduate School, CNRS, Laboratoire Charles Fabry, Palaiseau, France.

[5]Laboratoire Interdisciplinaire Carnot de Bourgogne, CNRS UMR 6303 Université de Bourgogne, 21000 Dijon, France.

*Corresponding author: mathieu.mivelle@sorbonne-universite.fr

ORCID:0000-0002-0648-7134

List of the main content:

Simulation parameters
Supporting figures S1 to S8

**Simulation parameters:**

The simulations carried out in this study were done by the finite difference time domain (FDTD) method performed on the commercial software Lumerical from Ansys. This method solves Maxwell's equations in space and time using a finite difference technique. Indeed, the FDTD method solves these equations on a discrete spatial and temporal grid. The dimensions of the 3D computational window for the simulations were 750x750x900 nm$^3$. The boundary conditions of this window are made of a perfectly matched layer (PML), avoiding any parasitic reflection inside the calculation window. Several meshes are used for the discretization of the computational space, a coarser non-uniform mesh of 4 to 16 nm for the external unstructured parts, of the simulation, a finer mesh of 4 nm for a central nanostructured part of 288x288x36 nm$^3$ in X, Y, and Z, respectively containing the nanostructures, and an even finer mesh of 1 nm for the part where the drift currents and magnetic field are calculated of 140x140x32 nm$^3$ in X, Y, and Z (Figure S1c). The choice of this mesh size for the central part is chosen because convergence in the amplitude of the magnetic field is observed starting from this mesh size. The excitation of the nanostructures is performed by a pulsed plane wave of duration 5.3 fs spectrally centered at a wavelength of 800 nm. The peak power of the pulse is $10^{10}$ W/cm$^2$, which corresponds to an energy slightly lower than 0.048 pJ applied to the plasmonic nanostructures (energy density of 53 µJ/cm$^2$). The convergence of the simulation was obtained when the energy inside the calculation window was lower than $10^{-5}$ of the initial injected energy. The textbook values of Johnson and Christy were used for the gold properties in these simulations.

Genetic Algorithm parameters: a binary matrix of 10x10 elements represents each structure. Half of the new structures in a new generation are produced by mutation, and the other half by breeding. Mutation was carried out at a rate of 10%, i.e., ten elements out of the 100 in a structure were changed randomly. Breeding was performed using two-point scattering breeding. In other words, two elements from two structures were randomly selected and interchanged from one structure to another.

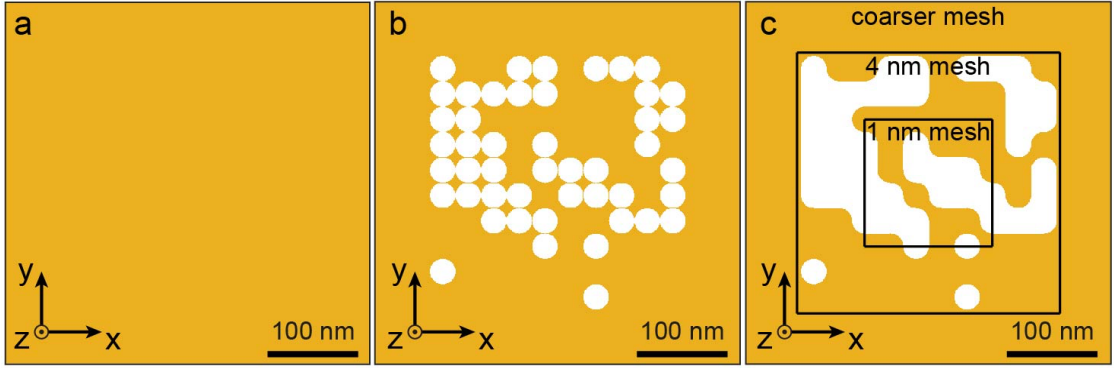

*Figure S1*. Construction of the elements constituting each generation of the genetic algorithm. Inside a) a 30 nm thick uniform layer of gold, b) holes are made according to a binary matrix acting as the DNA in the evolutionary process. c) The obtained structure is then smoothed to avoid all the roughnesses not experimentally realistic and generating non-physical effects. The different mesh areas are shown in c.

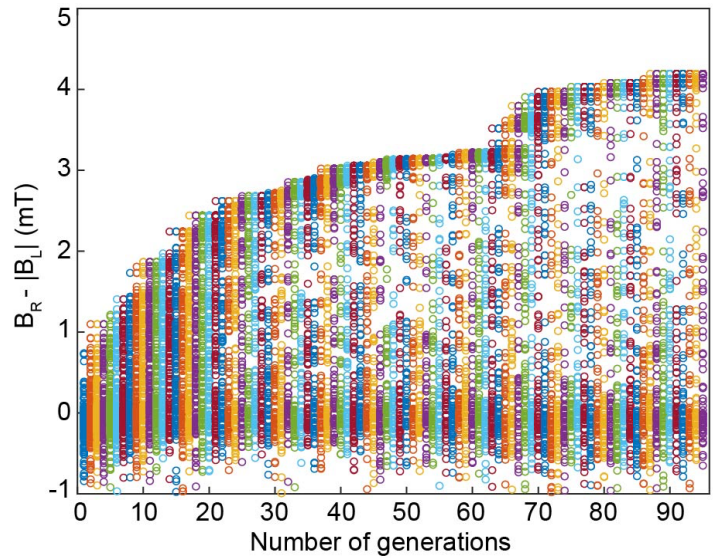

*Figure S2*. Evolutionary process. Evolution during the different generations of the optimization function that maximizes the difference $B_R$-abs($B_L$), with $B_R$ and $B_L$ being the magnetic fields **B** created by a right or left circular polarization, respectively. Each generation consists of 200 elements. The optimized structure appears after 87 generations.

Electronic and lattice temperature spatiotemporal evolution was carried out with a MATLAB finite elements model based on a two-temperature model, considering thermal conduction and out-of-equilibrium electron displacement[1].

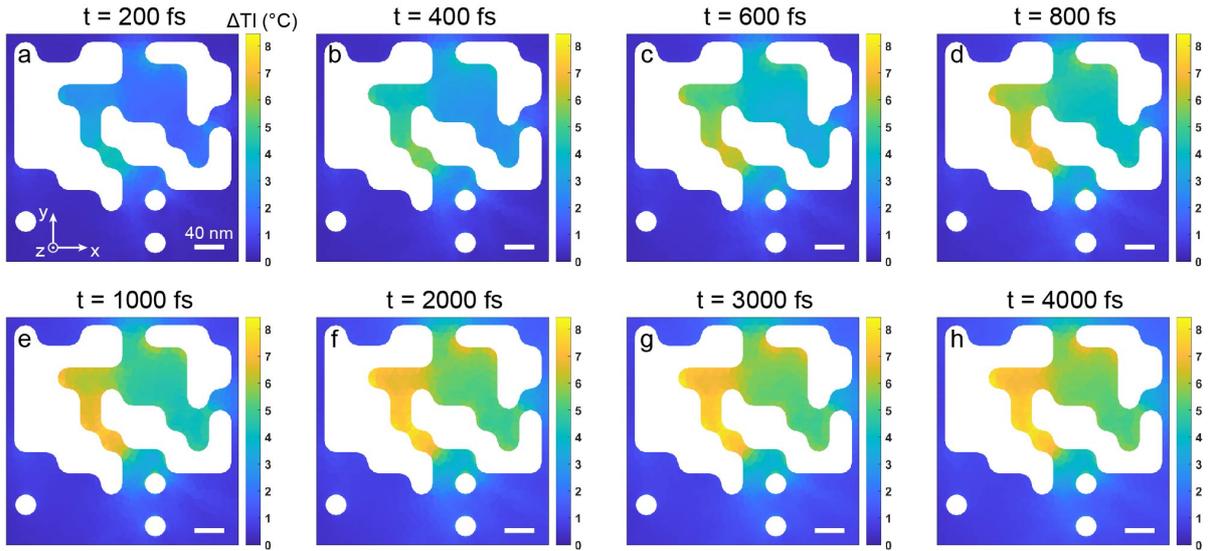

*Figure S3. Spatial distribution of lattice temperature generation in an XY plane at the Z center of the optimized nanostructure for different characteristic times after the optical pulse, a) 200 fs, b) 400 fs, c) 600 fs, d) 800 fs, e) 1000 fs, f) 2000 fs, g) 3000 fs, h) 4000 fs.*

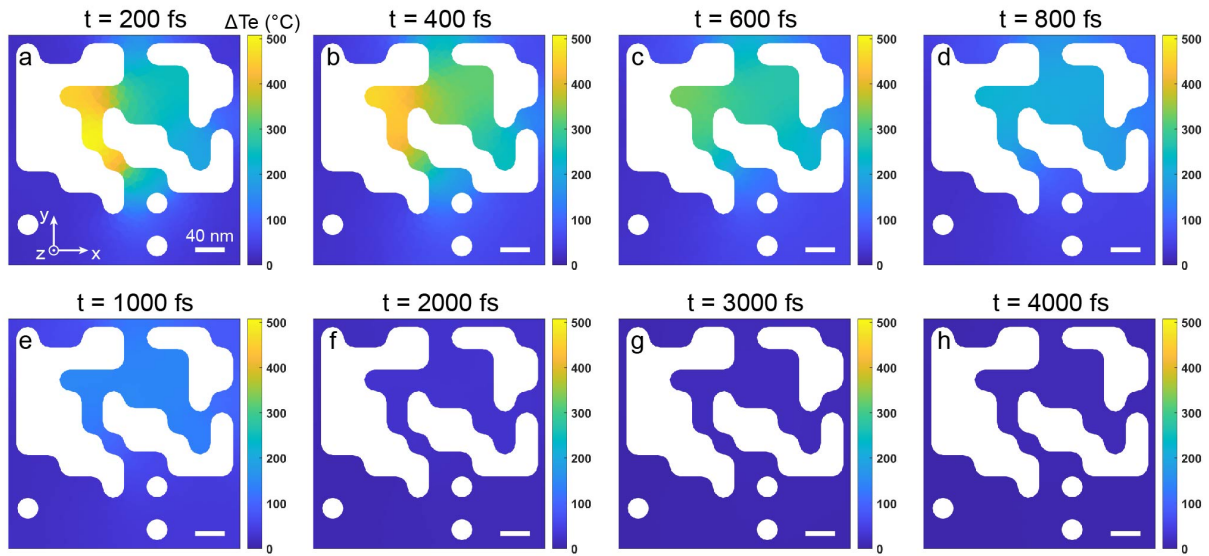

*Figure S4. Spatial distribution of electronic temperature generation in an XY plane at the gold surface of the optimized nanostructure for different characteristic times after the optical pulse, a) 200 fs, b) 400 fs, c) 600 fs, d) 800 fs, e) 1000 fs, f) 2000 fs, g) 3000 fs, h) 4000 fs.*

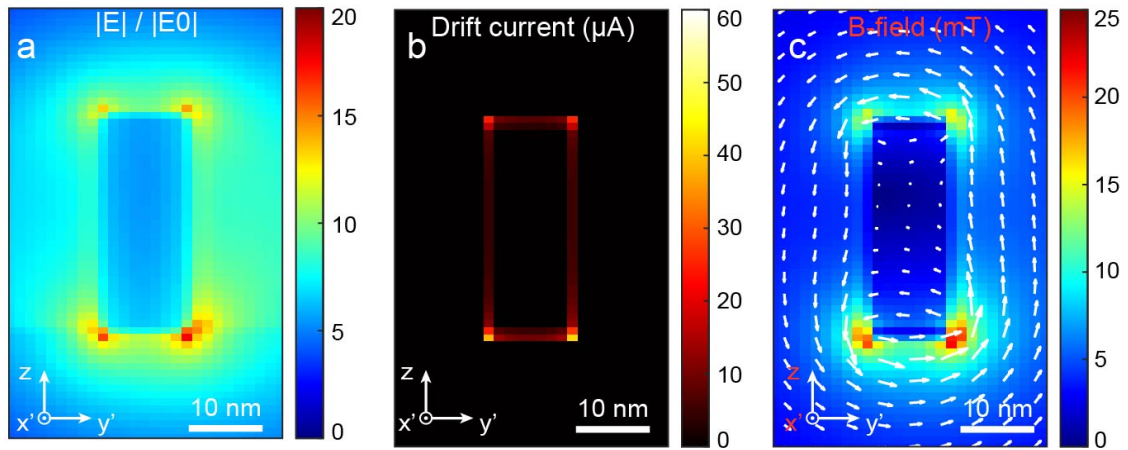

***Figure S5***. *Spatial distribution of the electric field enhancement in a Y'Z plane of the X'Y'Z coordinate, symbolized by the pink dotted line and shown in Figure 4a of the manuscript. Distribution b) of the drift currents described by equation 1) and c) of the resulting magnetic field in the same plane as a).*

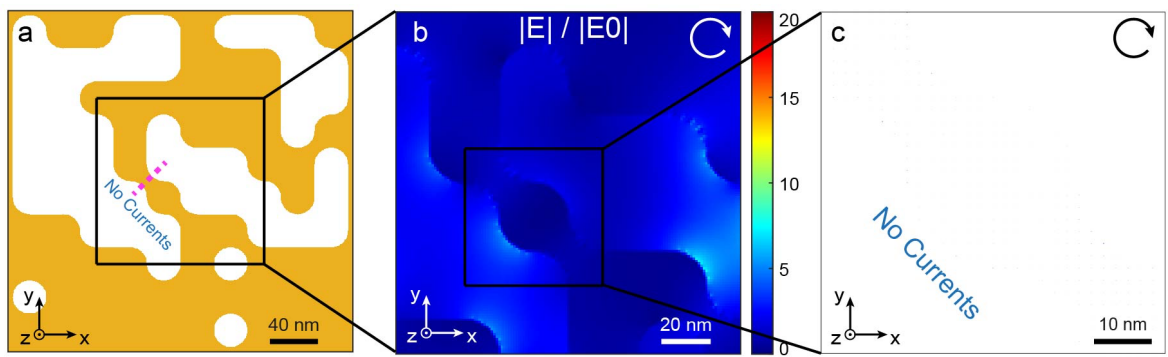

***Figure S6***. *Optical response of the optimized plasmonic nanostructure and associated drift currents for excitation by a left circular polarization. a) Schematic, in an XY plane, of the GA-optimized structure. b) Spatial distribution of the electric field enhancement at the Z-center of the structure shown in a) for an excitation by a left circular polarization. c) Spatial distribution of the drift currents generated (or rather not generated) by the electric field distribution shown in b).*

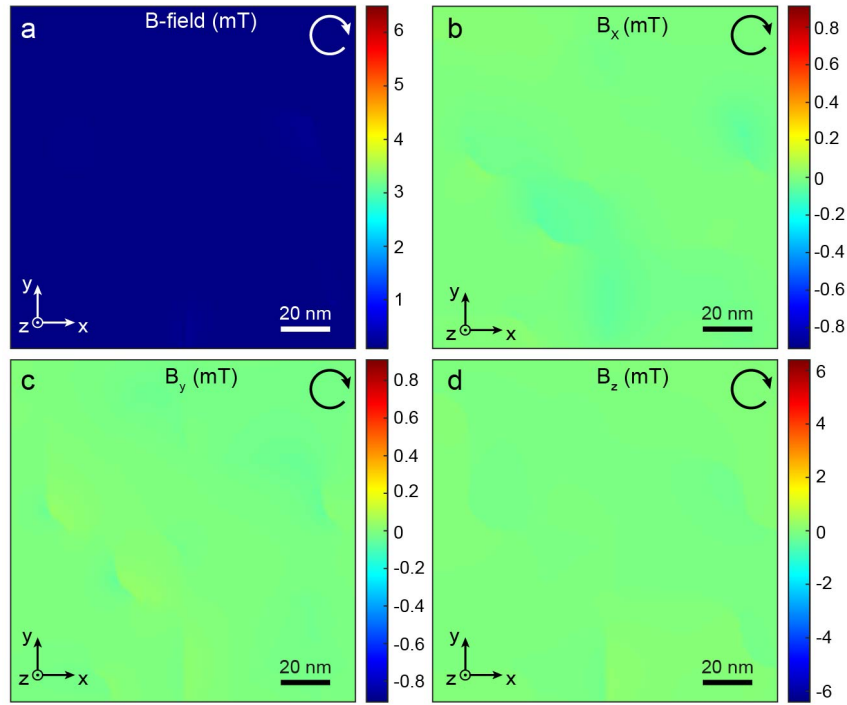

***Figure S7***. *Magnetic response of the optimized plasmonic nanostructure under left circular polarized excitation. Spatial distribution of the a) total magnetic field and its different components along b) X, c) Y and d) Z, generated at the Z center of the structure shown in Figure 2a of the manuscript.*

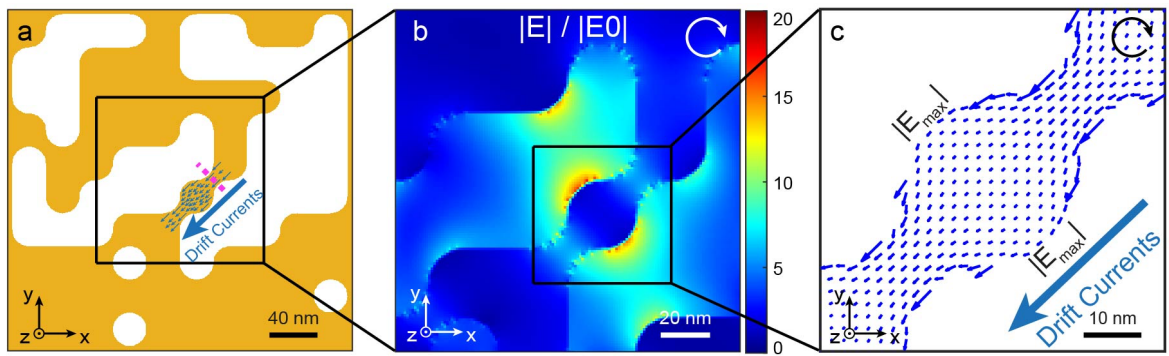

***Figure S8***. *Optical response of the mirror plasmonic structure. a) Schematic representation of the GA-optimized mirror antenna. b) Electric field enhancement at the Z-center of the mirror structure in the area enclosed by a black square in a). c) Drift currents created by the electric field distribution described in b) via equation 1 at the surface of the structure.*